\documentclass[%
 reprint,
 superscriptaddress,
 amsmath,amssymb,
 aps,
 hidelinks
]{revtex4-2}

\usepackage{graphicx}% Include figure files
\usepackage{dcolumn}% Align table columns on decimal point
\usepackage{bm}% bold math
\usepackage{pdfcomment}

\usepackage{xcolor}
\usepackage{siunitx}
\usepackage{wasysym}
\usepackage{relsize}
\usepackage{xcolor}
\usepackage{braket}
\usepackage{multirow}
\usepackage[normalem]{ulem}
\usepackage[T1]{fontenc}
\newcommand{\toremove}[1]{}

\begin{document}

\title{Robust Microwave-Optical Photon Conversion Using Cavity Modes Strongly Hybridized with a Color Center Ensemble} 

\author{M. Khalifa}
\affiliation{Stewart Blusson Quantum Matter Institute, University of British Columbia, Vancouver, BC, Canada.} \affiliation{Department of Electrical and Computer Engineering, University of British Columbia, Vancouver, BC, Canada.}
\author{P. S. Kirwin}
\affiliation{Stewart Blusson Quantum Matter Institute, University of British Columbia, Vancouver, BC, Canada.} \affiliation{Department of Electrical and Computer Engineering, University of British Columbia, Vancouver, BC, Canada.}
\author{Jeff F. Young}
\affiliation{Stewart Blusson Quantum Matter Institute, University of British Columbia, Vancouver, BC, Canada.} \affiliation{Department of Physics and Astronomy, University of British Columbia, Vancouver, BC, Canada.}
\author{J. Salfi}
\email{jsalfi@ece.ubc.ca}
\affiliation{Stewart Blusson Quantum Matter Institute, University of British Columbia, Vancouver, BC, Canada.} \affiliation{Department of Electrical and Computer Engineering, University of British Columbia, Vancouver, BC, Canada.} \affiliation{Department of Physics and Astronomy, University of British Columbia, Vancouver, BC, Canada.}

\date{\today}% It is always \today, today,

\begin{abstract}
A microwave-optical photon converter with high efficiency ($>50$~\%) and low added noise ($\ll 1$ photon) could enable the creation of scalable quantum networks where quantum information is distributed optically and processed in the microwave regime. However, integrated converters demonstrated to date lack sufficient co-operativity or are too lossy to provide the required performance. Here we propose a bi-directional microwave-optical converter employing an ensemble of spin-bearing color centers hosted within a high-Q Si photonic resonator and coupled magnetically to a high-Q superconducting microwave resonator. We develop a theory for microwave-optical conversion when the ensemble of centers is strongly hybridized with one or both cavities, and find a counterintuitive operating point where microwave and optical photons are tuned to bare center/cavity resonances. Compared to the perturbative coupling regime, we find a substantially enhanced nonlinearity, making it possible to obtain the required co-operativity with reduced pump- and center-induced losses, and improved robustness to optical inhomogeneous broadening. Taking color center and optical pump-induced losses into account in both the Si photonic and superconducting resonators, we find that $\sim 95~\%$ total efficiency and added noise $\ll 1$ quanta is possible at low (\qty{}{\micro\watt}) pump powers for both Er- and T-centers in Si. Our results open new pathways towards quantum networks using microwave-optical converters.

\end{abstract}

\maketitle

\section{\label{sec:Intro}Introduction}
Distributing quantum information across a quantum network \cite{wolf2007quantum, kimbleQuantumInternet2008} where nodes operate in the microwave domain \cite{barends2014superconducting, nguyen2022blueprint, hendrickx2021four, philips2022universal}, will inevitably require a coherent interface between microwave photons and optical photons. Recently, several coherent microwave-optical converter (MOC) devices have been proposed to build quantum networks \cite{barzanjeh2012reversible, williamson2014magneto, javerzac2016chip, li2017quantum, wu2020microwave, liu2021one,barnett2020theory}, and MOCs relying on electro-optical \cite{rueda2016efficient, hease2020bidirectional, fan2018superconducting, holzgrafe2020cavity, mckenna2020cryogenic, xu2021bidirectional, witmer2020silicon, sahu2022quantum, sahu2023entangling}, opto-electro-mechanical \cite{higginbotham2018harnessing, mirhosseini2020superconducting, forsch2020microwave, brubaker2022optomechanical, weaverIntegratedMicrowaveopticsInterface2024, meesalaNonclassicalMicrowaveOptical2024}, and magneto-optical \cite{hisatomi2016bidirectional, ihn2020coherent, zhu2020waveguide, puel2024enhancement, lekavicius2017transfer, fernandez2015coherent, fernandez2019cavity, bartholomew2020chip, rochman2023microwave} nonlinearities have been demonstrated. On-chip MOCs are particularly appealing due to their compact size, potential for low pump powers, potential for integration.

Despite significant experimental progress \cite{rueda2016efficient, hease2020bidirectional, fan2018superconducting, holzgrafe2020cavity, mckenna2020cryogenic, xu2021bidirectional, witmer2020silicon, sahu2022quantum, sahu2023entangling,higginbotham2018harnessing, mirhosseini2020superconducting, forsch2020microwave, brubaker2022optomechanical, weaverIntegratedMicrowaveopticsInterface2024, meesalaNonclassicalMicrowaveOptical2024,hisatomi2016bidirectional, ihn2020coherent, zhu2020waveguide, puel2024enhancement, lekavicius2017transfer, fernandez2015coherent, fernandez2019cavity, bartholomew2020chip, rochman2023microwave}, integrated MOCs demonstrated to date do not meet efficiency ($\eta>50\%$), and added noise ($N_{\rm add}\ll1$) criteria for a quantum channel \cite{wolf2007quantum}. Both figures of merit are degraded by loss and/or excitation within the microwave and optical subsystems and in nonlinear media providing the conversion \cite{han2021microwave}. Losses often arise from strong optical pumps needed to compensate weak nonlinearities, while noise can be added from excitations in strongly nonlinear media that are difficult to cool \cite{higginbotham2018harnessing, brubaker2022optomechanical}. Significant improvements in converter performance (ideally approaching $100~\%$ efficiency) from stronger nonlinearities, reduced loss and reduced noise are needed to unlock the potential of MOCs for quantum networks.

Spin-bearing, optically active centers in solids are a promising system for high-efficiency, low-noise conversion via the magnetic coupling of their spin transitions to microwave photons and coupling of their orbital transitions to optical photons\cite{williamson2014magneto,barnett2020theory,fernandez2015coherent, lekavicius2017transfer, fernandez2019cavity, xie2021characterization,bartholomew2020chip, rochman2023microwave, higginbottom2023memory}. To date, the regime of large ensemble-cavity detuning has been considered theoretically cavity-enhanced MOC with color center ensembles \cite{williamson2014magneto, barnett2020theory}, and rare-earth ion-based MOC devices operating in this regime have reached $\sim1$\% efficiencies \cite{rochman2023microwave,xie2024scalable}. Greater efficiency could be obtained through increased center-cavity couplings $g$ and center number $N_A$, and decreased detuning $\delta$ and cavity loss $\kappa$ \cite{williamson2014magneto}.  Each of these changes brings the system closer the strong ensemble-cavity coupling regime, $2g\sqrt{N_A} \gtrsim \delta/2, \kappa/2$, for both the optical and microwave cavities, but this regime has yet to be described theoretically. 

Here we generalize the theory of MOC to the case where one or both of the microwave and optical cavities is strongly coupled to the center ensemble, for both homogeneously and inhomogeneously broadened ensembles, including dressing by the optical pump. We find that high-efficiency conversion is possible when the microwave and optical signal photons are nearly resonant with states that resemble the vacuum Rabi-split eigenstates of the isolated microwave and optical subsystems. Counterintuitively, we also find that near-unity conversion efficiency is possible when the optical and microwave signal photons are close to the bare center/cavity resonances for strong ensemble-cavity coupling. This strong coupling regime allows us to maintain high co-operativity and conversion efficiency at lower pump power, and benefits from pump dressing to suppress negative effects of inhomogeneous broadening while retaining low center-induced losses.

\begin{figure*}
    \centering
    \includegraphics[width=0.98\linewidth]{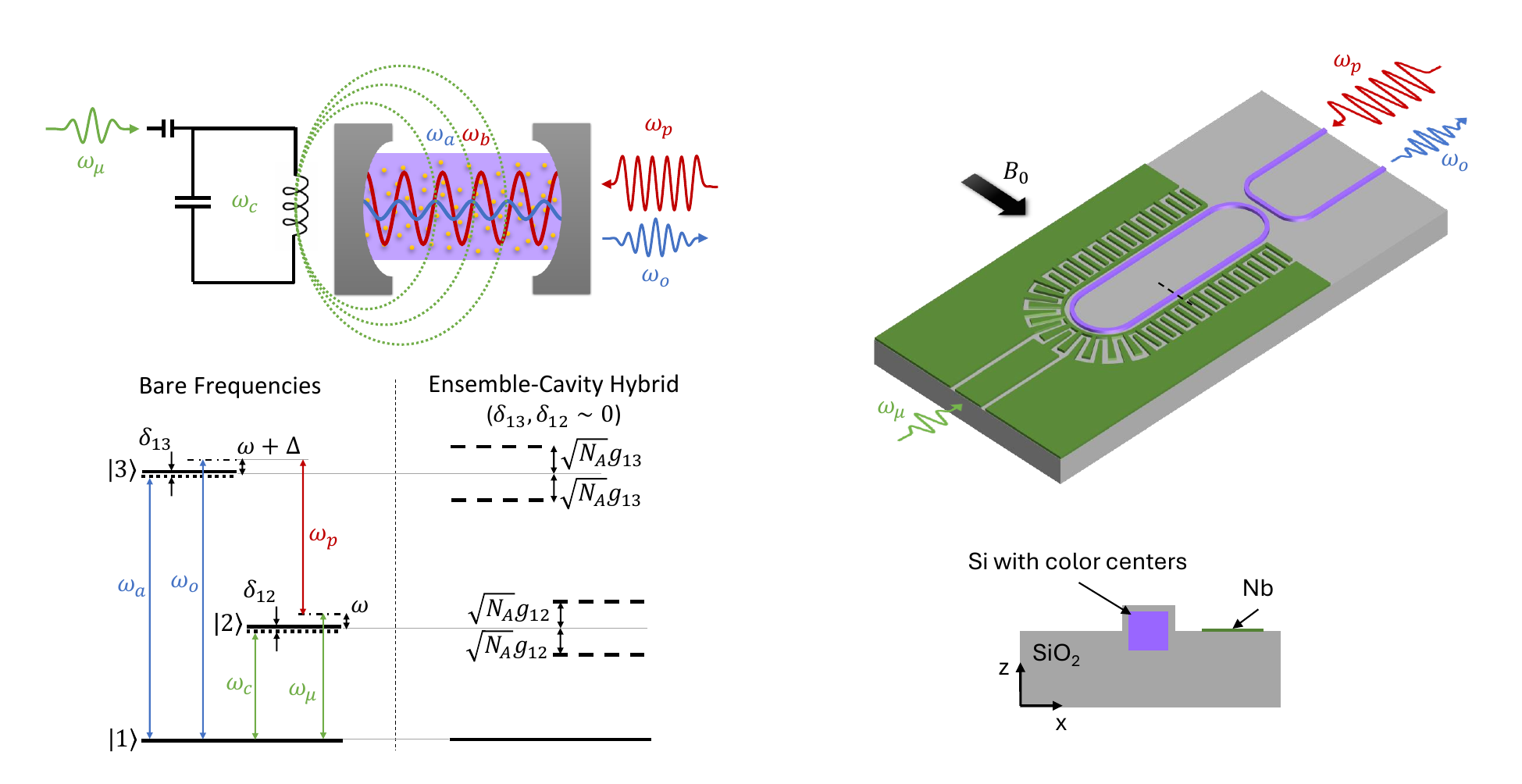}
    \put(-490,230){(a)}
    \put(-490,140){(b)}
    \put(-220,230){(c)}
    \put(-220,75){(d)}
    \caption{
    Si-integrated microwave-optical converter. (a) Schematic of color centers in Si coupled optically to a Si cavity with signal and pump resonances and magnetically to a superconducting microwave cavity. The conversion between the microwave and optical photons at $\omega_\mu$ and $\omega_o$ respectively is induced by an optical pump at $\omega_p$. (b) Energy diagram for color center ensemble and cavity resonances $\omega_a$ and $\omega_c$ for (left) no ensemble-cavity coupling and (right) ensemble-cavity coupled system at zero detunings with pump off. The color center ensemble is assumed to be homogeneously broadened in the diagram. (c) Device schematic comprising an optical racetrack cavity and a distributed superconducting microwave cavity loaded with an interdigitated capacitor, with applied magnetic field. (d) Cross-section view of the device at the dashed line in the main panel.}
    \label{fig:dev}
\end{figure*}
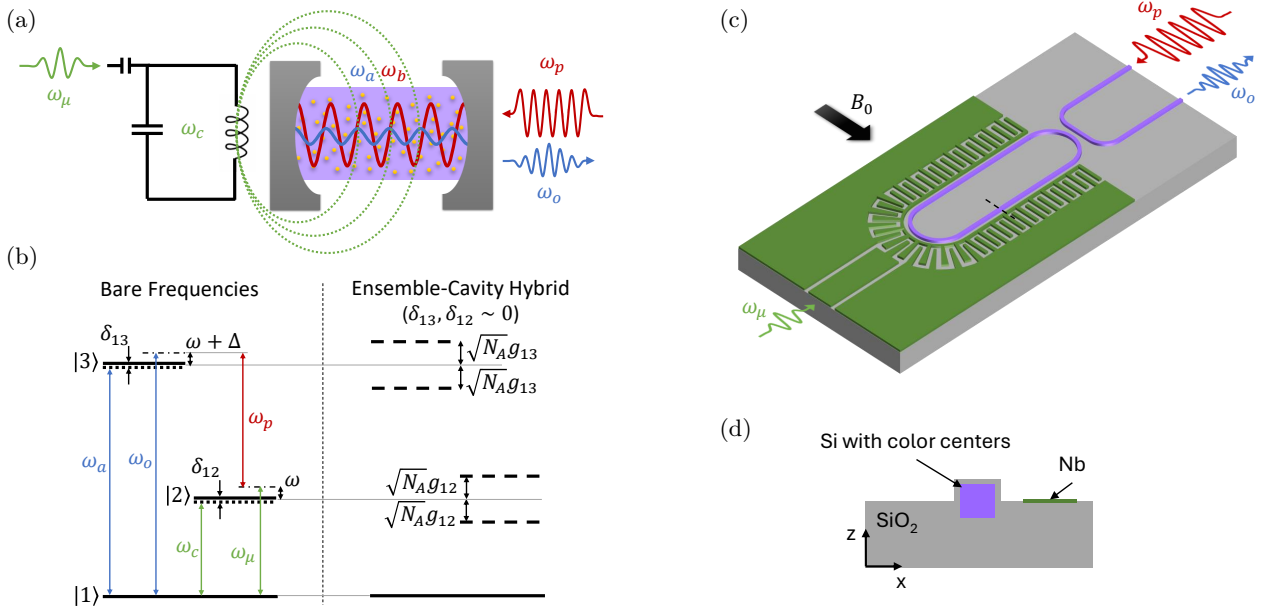

To showcase the potential of microwave-optical conversion in the strong ensemble-cavity coupling regime, we propose a device consisting of a high-Q Si photonic racetrack resonator containing color centers and a high-Q superconducting resonator. We predict the MOC efficiency and noise for optical and microwave photons tuned to the bare center and cavity resonances, considering the following nonidealities: (1) color center-induced loss and dispersion, (2) two-photon and free-carrier absorption loss in Si, (3) optical loss and optically-induced microwave loss to the superconducting resonator, and (4) realistic color center inhomogeneous broadening. The proposed device benefits from strong two-dimensional optical confinement in Si resonators to limit losses due to superconducting quasiparticle generation. The superconducting resonator is loaded with an interdigitated capacitor \cite{malnou2021three}, which serves to match its length to the photonic resonator and reduce its impedance, both of which enhance nonlinearity. Benefiting from long spin lifetimes and narrow spin linewidths in Si \cite{tyryshkin2012electron, saeedi2013room, kobayashi2021engineering,bergeron2020silicon,macquarrie2021generating}, we find a concentration of $\sim 10^{16}$~cm$^{-3}$ T-centers ($\sim 10^{17}$ Er-centers) \cite{bergeron2020silicon, macquarrie2021generating}, with optical inhomogeneous broadening up to 1 GHz (similar to reported values) provides near-unity efficiency ($\sim 95\%$) with added noise $\ll 1$~photon at pump powers $\lesssim \qty{1}{\micro\watt}$.

\section{Concept and Model}

The proposed MOC (Fig.~\ref{fig:dev}(a)) employs three-wave mixing (3WM) between a microwave signal, optical signal, and optical pump having photon frequencies $\omega_\mu$, $\omega_o$, and $\omega_p=\omega_o-\omega_\mu$, respectively. The 3WM process is resonantly enhanced by microwave cavity, optical cavity, and optical pump cavity modes with bare frequencies $\omega_c$, $\omega_a$, and $\omega_b$, respectively \cite{tsang2010cavity,tsang2011cavity}. The three cavity modes overlap spatially inside a medium hosting an ensemble of color centers with microwave and optical transition frequencies $\omega_{12}$, $\omega_{13}$, and $\omega_{23}$, which provides the 3WM nonlinearity. 

The Hamiltonian describing the system in the frame rotating at the resonant frequencies $\omega_c$ and $\omega_a$ is \cite{williamson2014magneto, fernandez2019cavity}
\begin{equation}\label{eqn:Ham}
\begin{split}
    \hat{H}/\hbar & = \sum_k \left( \delta_{13,k} \hat{\sigma}_{33,k} + \delta_{12,k} \hat{\sigma}_{22,k} \right) \\
    & + \sum_k \left( g_{13,k} \hat{\sigma}_{31,k} \hat{a} + g_{12,k} \hat{\sigma}_{21,k} \hat{c} \right) + h.c. \\
    & + \sum_k \left( \Omega_{p,k} e^{-i\Delta t} \hat{\sigma}_{32,k} \right) + h.c.,
\end{split}
\end{equation}
where $\hat{\sigma}_{ij,k}$ is a Pauli operator for the $k$th color center, with $i,j\in \{1,2,3\}$, $\hat{a}$ ($\hat{c}$) is the annihilation operator of the optical (microwave) signal photon. Here, \mbox{$\delta_{13,k}=\omega_{13,k}-\omega_a$} and \mbox{$\delta_{12,k}=\omega_{12,k}-\omega_c$} are the detunings of the orbital and spin transitions of the $k$th center from the bare optical and microwave cavity frequencies, respectively, which may vary from center to center. Additionally, $g_{13,k}$, $g_{12,k}$ are the optical and microwave center-cavity coupling rates, and $\Omega_{p,k}$ is the pump Rabi frequency for the $\ket{2} \leftrightarrow \ket{3}$ optical transition (Fig.~\ref{fig:dev}(b)), which may vary between centers due to spatial inhomogeneity of the cavity modes. $\Delta=\omega_p - (\omega_a-\omega_c)$ is the shift of the pump frequency from the difference between the bare optical and microwave cavity frequencies. 

We solve the Heisenberg equations of motion for Eq.~(\ref{eqn:Ham}) with single-mode optical and microwave reservoir channels and center dampings. The conversion efficiency is predicted from input-output theory, taking into account internal loss and port reflection. Our treatment differs from past works \cite{williamson2014magneto,fernandez2019cavity,barnett2020theory} by allowing arbitrary center-cavity detunings and time-varying cavity fields $\{\hat{a},\hat{c}\}$ and color center polarizations $\{\hat{\sigma}_{13},\hat{\sigma}_{12}\}$ relevant in the hybrid ensemble-cavity regime. In our analytic treatment we neglect upper state populations and nonlinear operator terms in the Heisenberg equations of motion \cite{williamson2014magneto,barnett2020theory}, and validate these assumptions using a numerical solution presented in the Supplementary Information. 

We find that the bidirectional total conversion efficiency is given by

\begin{equation}\label{eqn:full_eta}
    \eta = \frac{\kappa_a^{\rm ex} \kappa_c^{\rm ex} |\xi_{ac}|^2}{\left| \xi_{ac}^2 + \left[ \kappa_a/2 - i(\omega+\Delta-\xi_a) \right] \left[ \kappa_c/2 - i(\omega-\xi_c) \right] \right|^2},
\end{equation}
where $\kappa_{a/c}^{\rm ex}$ ($\kappa_{a/c}\ge \kappa_{a/c}^{\rm ex}$) is the coupling (total) loss of the mode $a/c$, not including color center losses. The center losses depend on the center relaxation rates $\gamma_{13,k}$ and $\gamma_{12,k}$ for optical and microwave excitations. Furthermore, 
\begin{subequations}\label{eqn:xi}
    \begin{equation}
        \xi_a = \sum_{k=1}^{N_A} \frac{g_{13,k}^2 (\omega - \Delta_{12,k})}{(\omega + \Delta - \Delta_{13,k})(\omega - \Delta_{12,k}) - \Omega_{p,k}^2},
    \end{equation}
    \begin{equation}
        \xi_c = \sum_{k=1}^{N_A} \frac{g_{12,k}^2 (\omega + \Delta - \Delta_{13,k})}{(\omega + \Delta - \Delta_{13,k})(\omega - \Delta_{12,k}) - \Omega_{p,k}^2},
    \end{equation}
    \begin{equation}
        \xi_{ac} = \sum_{k=1}^{N_A} \frac{g_{12,k} g_{13,k} \Omega_{p,k}}{(\omega + \Delta - \Delta_{13,k})(\omega - \Delta_{12,k}) - \Omega_{p,k}^2},
    \end{equation}
\end{subequations}
where $\omega=\omega_{\mu} - \omega_c$ and $\Delta_{13,k} = \delta_{13,k}-i\gamma_{13,k}/2$ and $\Delta_{12,k} = \delta_{12,k}-i\gamma_{12,k}/2$ are the complex optical and microwave cavity-center detunings.

Under the assumption that the dominant source of added noise is thermal noise in the microwave cavity, the input-referred added noise for microwave-optical and optical-microwave conversion are given respectively by
\begin{subequations}\label{eqn:nois}
\begin{equation}
    N_{\rm MO} = \frac{\kappa_c^{\rm in}}{\kappa_c^{\rm ex}} N_{th},
\end{equation}
\begin{equation}
    N_{\rm OM} = \frac{|\kappa_a/2-i(\omega+\Delta-\xi_a)|^2}{|\xi_{ac}|^2} \frac{\kappa_c^{\rm in}}{\kappa_a^{\rm ex}} N_{th},
\end{equation}
\end{subequations}
where $N_{th}=[\exp{(\hbar\omega_c/k_BT)}-1]^{-1}$ is the microwave thermal noise population, and $\kappa_c^{\rm in}$ is the microwave cavity intrinsic loss (see Supplementary Information). 

\section{Analysis and Discussion}
Here we discuss the hybrid ensemble-cavity converter and its performance. First, we use our model to quantify device performance as a function of the cavity and photon frequencies (relative to centers), pump strength, color center density, vacuum Rabi coupling, and external coupling $\kappa^{\rm ex}$, considering only color center loss. Next, we show that evanescent optical confinement in the Si waveguide (Fig.~\ref{fig:dev}(d)) enables exponential suppression of pump photon loss to the superconductor with only small changes to internal conversion efficiency, and calculate pump-dependent intrinsic nonlinear losses in Si, and losses in the optical and microwave modes due to optically generated quasiparticles in the superconducting cavity. Finally, we predict efficiency and added noise for both T-centers and Er-centers, considering all losses and inhomogeneous broadening.

\subsection{Efficiency and center-induced losses}\label{sec:etain}

In this section we calculate conversion efficiency using our model, considering color center absorption/relaxation as the only source of loss. Three categories of model parameters are considered; (A) the center-cavity detunings $\delta_{13/12}$ and photon-cavity detunings $\omega$ and $\Delta$, (B) the color center number $N_A$, the cavities external coupling rates $\kappa_{a/c}^{\rm ex}$, and the number of pump photons $N_p$, which have practical limits, and (C) color center and material properties (e.g. dipole moments, decay rates, intrinsic cavity losses, inhomogeneous distributions, etc.)  For simplicity, we limit calculations in this section to homogeneously broadened ensembles, though we qualitatively discuss inhomogeneous broadening. Calculations with all loss mechanisms and inhomogeneous broadening are presented in Section~\ref{sec:eta}. For simplicity, we assume spatially uniform bosonic fields and drop $k$ subscripts on $\gamma_{ij}$, $g_{ij}$, and $\Omega_p$. We consider $\delta_{12} = \delta_{13} = 0$, opposite from the condition $\delta_{12/13}\gg g_{12/13}\sqrt{N_A}$ for weak hybridization \cite{williamson2014magneto,barnett2020theory}.

\begin{figure*}
    \centering
    \includegraphics[width=0.99\linewidth]{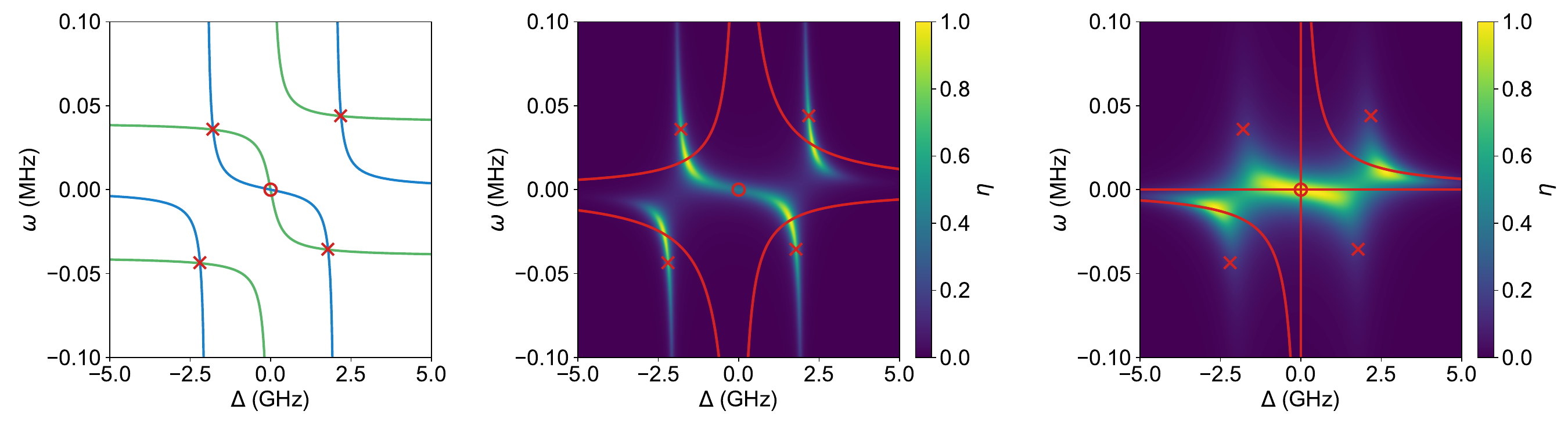}
    \put(-500,140){(a)}
    \put(-350,140){(b)}
    \put(-170,140){(c)}
    \caption{Conversion efficiency spectra at zero center-cavity detunings \mbox{($\delta_{13}=\delta_{12}=0$)}. (a) Contour lines of \mbox{$\omega+\Delta-\text{Re}\{\xi_a\}=0$} (blue) and \mbox{$\omega-\text{Re}\{\xi_c\}=0$} (green) at the T-center parameters from Table \ref{tab:params}. The four red crosses denote the positions of the near-resonance intersection points of the blue and green lines, while the red circle denotes the fifth intersection point at the origin arising from the nonlinear coupling between the microwave and optical systems at finite $\Omega_p$. (b) Conversion efficiency spectrum for the T-center parameters from Table \ref{tab:params} except for $\kappa_a=\qty{0.25}{GHz}$. The red lines are the contour lines of the matching condition \mbox{$4|\xi_{ac}|^2=\kappa_a \kappa_c$}. 
    The red crosses and circle denote the same points as in (a). 
    (c) Same as panel (b) but with $\kappa_a=2$ GHz, such that matching is obtained at $\omega=\Delta=0$.
    }
    \label{fig:chi}
\end{figure*}

{\it Homogeneous Limit: }Examining Eq.~(\ref{eqn:full_eta}) with centers and cavities on resonance ($\delta_{12} = \delta_{13} = 0$), where previous treatments do not apply, reveals that unity efficiency is obtained for optical and microwave photon frequencies defined by the intersection of optical dispersion contours \mbox{$\omega+\Delta-\text{Re}\{\xi_a\} = 0$} (condition (i)) and microwave dispersion contours \mbox{$\omega-\text{Re}\{\xi_c\} = 0$} (condition (ii)), when a matching condition \mbox{$C=4|\xi_{ac}|^2/\kappa_a\kappa_c=1$} for the magneto-optic cooperativity $C$ is satisfied. The optical and microwave contours and their five intersection points are shown in Fig.~\ref{fig:chi}(a) for the parameters in Table \ref{tab:params} that we consider for T-centers.  For large $|\omega|$ ($|\Delta|$), the microwave (optical) dispersion contours approach lines representing vacuum Rabi-split states of the microwave (optical) subsystems in the absence of a pump, with energies $\omega = \pm \sqrt{N_A}g_{12}$ ($\omega+\Delta = \pm \sqrt{N_A}g_{13}$). 
More practical conditions for high efficiency that take into account cavity-reservoir coupling and center loss are $|\omega+\Delta-\xi_a| \ll \kappa_a$ and $|\omega-\xi_c| \ll \kappa_c$. 

The conversion efficiency is plotted in Fig.~\ref{fig:chi}(b) for the same parameters used in Fig.~\ref{fig:chi}(a). Cavity couplings are given in Table~\ref{tab:params}. From Fig.~\ref{fig:chi}(b), we see that the efficiency is highest at the intersections of the optical contour and the matching contour. This is because the microwave reservoir coupling $\kappa_c$ is larger than the microwave cavity vacuum Rabi splitting, such that $|\omega-\text{Re}\{\xi_c\}| \ll \kappa_c$ applies throughout Fig.~\ref{fig:chi}(b), while the optical reservoir coupling $\kappa_a$ is smaller than the optical cavity vacuum Rabi splitting. If both cavities operate in the strong coupling limit (not shown), the highest efficiencies in Fig.~\ref{fig:chi}(b) will occur near where the $C=1$ contours overlap with the four crosses. We also see that appreciable efficiencies are obtained as $\omega+\Delta$ approaches $\pm \sqrt{N_A}g_{13}$, and at the circular point defined by the origin $\Delta=\omega=0$. The latter operation point is counterintuitive because it does not occur near center-microwave-cavity and center-optical-cavity hybridized states, and is further discussed in Supplementary Information S1.G. Both signal photon frequencies are equal to the respective {\it bare} cavity and center resonance frequencies, {\it substantially off-resonance} from the vacuum Rabi-split center-optical-cavity hybrid states. 

The conversion efficiency can be optimized at the operating point with photons at the bare frequencies ($\Delta=\omega=0$) by choosing parameters to enable matching ($C=1$) at this point. A map of the efficiency is plotted in Fig.~\ref{fig:chi}(c) with $\kappa_a^{\rm ex}$ increased to \qty{2}{GHz}, while leaving other parameters fixed, to obtain $C=1$ at the origin, which occurs when
\begin{equation}
\label{eqn:cond1}
\Omega_p = \frac{2N_Ag_{12}g_{13}}{\sqrt{\kappa_a\kappa_c}}.
\end{equation}

Notably, the co-operativity is enhanced significantly in the strong ensemble-cavity coupling case compared to the weak coupling case. For the latter, \mbox{$\delta_{1i}\gg g_{1i}\sqrt{N_A}$} is needed to avoid hybridization of individual cavity photons to the ensemble, and \mbox{$\xi_{ac}=N_Ag_{12}g_{13}\Omega_p/\left(\delta_{12}\delta_{13}-\Omega_p^2\right)$}. Smaller detunings can be considered for the strong ensemble-cavity coupled case, where for \mbox{$\delta_{12}=\delta_{13}=0$} and \mbox{$\omega=\Delta=0$}, \mbox{$\xi_{ac}=N_A g_{12}g_{13}\Omega_p/\left(\gamma_{12}\gamma_{13}-\Omega_p^2\right)$}, provided that \mbox{$\Omega_p^2\gg \gamma_{12}\gamma_{13}$} to ensure low loss. Using the inequalities \mbox{$\delta_{1i}\gg g_{1i}\sqrt{N_A}$}, we see that co-operativity is enhanced by \mbox{$\left(g_{12}g_{13}N_A/\Omega_p^2-1\right)^2 \gg 12$ ($16$)} for \mbox{T- (Er-)} center parameters in Table \ref{tab:params}. For example, with \mbox{$\delta_{1i}= 4g_{1i}\sqrt{N_A}$}, the co-operativity enhancement is $300$ ($400$) for T- (Er-) centers. This enhancement is important because it allows us to increase the co-operativity with fixed center density and pump power (Rabi frequency) by reducing detunings to the strong coupling regime. 

\begin{table}
\caption{Parameters considered for the T- and Er-centers.}
    \label{tab:params} 
    \centering
    \begin{tabular}{|l|l|l|l|}
        \cline{2-4}
        \multicolumn{1}{l|}{}&Parameter & T-center & Er-center \\
        \hline
        \multirow{5}{*}{(B)}
        & $\Omega_p$ & \qty{4}{MHz} & \qty{4.5}{MHz} \\
        & $N_A$ & $10^6$ & $10^7$ \\
        & $\kappa^{\rm ex}_c$ & \qty{0.8}{MHz} & \qty{0.8}{MHz} \\
        & $\kappa^{\rm ex}_a$ [Fig.~\ref{fig:chi}(b)] & \qty{0.25}{GHz} & \multicolumn{1}{c|}{-} \\
        & $\kappa^{\rm ex}_a$ [Fig.~\ref{fig:chi}(c), Fig.~\ref{fig:eta}] & \qty{2}{GHz} & \qty{2}{GHz} \\
        \hline
        \multirow{4}{*}{(C)}
        & $g_{12}$ & \qty{40}{Hz} & \qty{0.3}{kHz} \\
        & $g_{13}$ & \qty{2}{MHz} & \qty{30}{kHz} \\
        & $\gamma_{12}$ & \qty{1}{Hz} & \qty{1}{kHz} \\
        & $\gamma_{13}$ & \qty{1}{MHz} & \qty{10}{kHz} \\
        \hline
    \end{tabular} 
\end{table}

{\it Inhomogenous broadening:} A key advantage of the operating point \mbox{$\delta_{12}=\delta_{13}=\omega=\Delta=0$} appears when considering the inhomogeneous broadening of the centers.  The Supplementary Information describes how the efficiency is rigorously calculated when including a two dimensional uncorrelated Gaussian distribution of widths $\sigma_{13}$ and $\sigma_{12}$ for the optical and microwave detunings centered around mean values $\left\langle\delta_{13}\right\rangle$ and $\left\langle\delta_{12}\right\rangle$. Operating at zero detunings (\mbox{$\left\langle\delta_{12}\right\rangle=\left\langle\delta_{13}\right\rangle=\omega=\Delta=0$}) makes $\xi_a$ and $\xi_c$ antisymmetric functions of $\delta_{12}$ and $\delta_{13}$, respectively (see Eq.~(\ref{eqn:xi}a,b) and Fig. S2). Therefore, the summation over the inhomogeneously broadened $\delta_{12/13}$ yields $\text{Re}\{\xi_a\}=\text{Re}\{\xi_c\}=0$, which fulfills conditions (i) and (ii) at $\omega=\Delta=0$. We therefore obtain enhanced robustness to inhomogeneous broadening by operating at zero detunings.

Another condition that assists in maintaining high efficiency for inhomogeneous ensembles is \mbox{$\Omega_p \gg \sqrt{\sigma_{12}\sigma_{13}}$}. If \mbox{$\Omega_p \ll \sqrt{\sigma_{12}\sigma_{13}}$}, $\xi_{ac}$ and the co-operativity are reduced. This is because terms with positive and negative $\delta_{12/13,k}$ cancel each other since $\xi_{ac}$ is an odd function of $\delta_{12/13,k}$.
However, when $\Omega_p \gg \sqrt{\sigma_{12}\sigma_{13}}$, $\xi_{ac}$ is almost independent of $\delta_{12/13,k}$. This condition extends the region where the magneto-optic coupling $\xi_{ac}$ is independent of $\delta_{12}$ and $\delta_{13}$ beyond the inhomogeneous broadenings.

{\it Color center loss:} The color centers that are responsible for microwave-optical conversion also introduce undesirable losses $|2\text{Im}\{\xi_{a/c}\}|$. 
For a strong pump, the color center loss can easily be evaluated, and the condition of high external quantum efficiency where reservoir coupling exceeds the loss simplifies to \mbox{$\kappa_a \gg N_Ag_{13}^2\gamma_{12}/\Omega_p^2$} and \mbox{$\kappa_c \gg N_Ag_{12}^2\gamma_{13}/\Omega_p^2$}. These conditions set practical bounds on $N_A$, i.e., the color center density in the host material, for given pump and cavity couplings. 

\subsection{Filling factor and cavities losses}
\label{sec:ffloss}
High quantum efficiency in the proposed device relies on appreciable spatial overlap of the color centers to the microwave signal, optical pump, and optical signal modes (quantified by filling factors) and low losses. The proposed device is shown in Fig.~\ref{fig:dev}(c,d). Here, the optical signal and pump are taken to be quasi-transverse electric (TE-like) and quasi-transverse magnetic (TM-like) modes, respectively, though a design based on coupled racetracks with two TE-like modes is also possible \cite{holzgrafe2020cavity, mckenna2020cryogenic}. The microwave cavity is a distributed superconducting resonator loaded with an interdigitated capacitor (IDC). The IDC loading increases the filling factor by reducing the length of the microwave cavity to match the optical cavity, reduces the characteristic impedance \cite{malnou2021three, parker2022degenerate}, and enhances the zero-point magnetic fluctuations \cite{haikka2017proposal, pla2018strain}. We consider an Nb resonator, which is robust to Tesla-level magnetic fields \cite{kroll2019magnetic}, though NbTiN is a viable alternative \cite{samkharadze2016high, khalifa2023nonlinearity}. After calculating the filling factor, we determine the optical and microwave losses due to absorption of optical photons in the superconductor, which generates quasiparticles, in addition to the Si two-photon and free-carrier absorption losses. We use these results to predict device performance in Sec.~\ref{sec:eta}. 

\begin{figure}
    \centering
    \includegraphics[width=0.99\columnwidth]{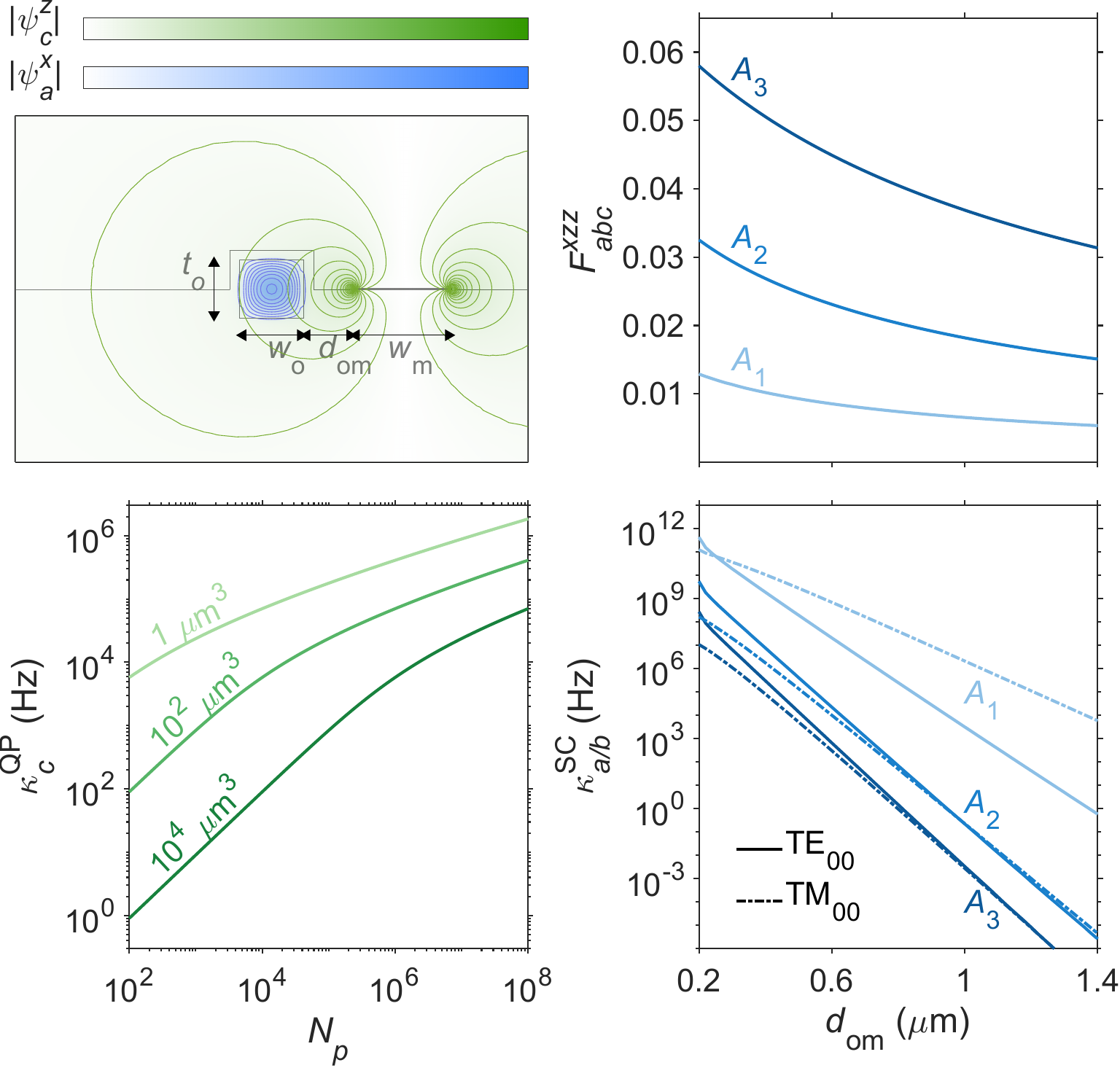}
    \put(-242,237){(a)}
    \put(-122,237){(b)}
    \put(-122,122){(c)}
    \put(-242,122){(d)}
    \caption{Filling factor, optical loss rates to superconductor, and microwave loss rate. (a) Calculated contours of TE\textsubscript{00} mode $a$ electric field along $x$ and microwave mode $c$ magnetic field along $z$. (b) Filling factor versus $d_{\rm om}$ for \mbox{$A_1=0.4\times\qty{0.22}{\micro\meter^2}$}, \mbox{$A_2=0.7\times\qty{0.5}{\micro\meter^2}$}, and \mbox{$A_3=1.3\times\qty{1.2}{\micro\meter^2}$} \mbox{($A=w_{\rm o} \times t_{\rm o}$)}. (c) Optical mode loss rates due to optical absorption in the superconductor, versus $d_{\rm om}$, for the same geometries. 
    The superconducting resonator geometry is \mbox{$t_{\rm m}=\qty{20}{nm}$} and \mbox{$w_{\rm m}=\qty{2}{\micro\meter}$}, and refractive index \mbox{$\approx 1+7i$} for Niobium at \qty{1300}{nm} and \qty{4}{K} \cite{golovashkin1969optical}. (d) Quasiparticle loss rate of the microwave mode versus optical pump photons number at $\kappa_{b}^{\rm SC}=\qty{300}{Hz}$ for three different superconductor volumes. 
    }
    \label{fig:geo}
\end{figure}

We begin by calculating the filling factor \mbox{$F^{ijk}_{abc}=|\int_{V_{\rm o}} \psi_a^i \psi_b^j \psi_c^k \,dV|/V_{\rm o}$} for the unitless mode fields $\psi_m^i$, for $m \in \{a,b,c\}$ and field directions $i,j,k \in \{x,y,z\}$ (see the Supplementary Information for details). We assume a fixed resonator width $w_{\rm m}=\qty{2}{\micro\meter}$. Fig.~\ref{fig:geo}(a) shows example contours for the TE\textsubscript{00} optical signal mode electric field $\psi_a^x$ along $x$ and the microwave mode magnetic field $\psi_c^z$ along $z$ at the narrow part of the IDC. The filling factor $F_{abc}^{xzz}$ is shown in Fig.~\ref{fig:geo}(b) as a function of distance $d_{\rm om}$ between the optical and the microwave cavities, for different waveguide cross-sectional areas ($w_{\rm o}, t_{\rm o}$). We see that (i) as $d_{\rm om}$ increases, $F$ only modestly decreases, and that (ii) larger optical waveguide cross-sections yield larger filling factors, effectively capturing more microwave magnetic field. 

We next evaluate the optical losses. Despite the weak dependence of the filling factor on distance, the loss rates of optical photons to the superconductor $\kappa_{a,b}^{\rm SC}$ reduce exponentially with distance $d_{\rm om}$, as shown in Fig.~\ref{fig:geo}(c). Indeed, $\kappa_{a,b}^{\rm SC}$ decreases exponentially by ten orders of magnitude as $d_{\rm om}$ increases from $\qty{0.2}{\micro\meter}$ to $\qty{1.4}{\micro\meter}$, although $F$ decreases modestly (by a factor of $\sim 2$) over the same range (Fig.~\ref{fig:geo}(b)). Furthermore, larger optical cross-sections reduce the evanescent penetration of the pump mode into the superconductor. Beyond quasiparticle generation in the superconductor and color center loss, we also consider nonlinear optical loss in Si from two-photon absorption (TPA) and free carrier absorption (FCA). Detailed calculations for these mechanisms are presented in the Supplementary Information. We show that when these loss mechanisms become appreciable, they reduce external quantum efficiency and cause $N_p$ to saturate with pump power, which is unfavourable for efficiency and noise. Smaller mode volumes enhance co-operativity but increase nonlinear loss by enhancing fields. 

Next we calculate the loss rate of the microwave photons in the superconductor $\kappa_c^{\rm QP}$ due to excess quasiparticles, an important loss mechanism in MOCs \cite{fan2018superconducting,mckenna2020cryogenic,holzgrafe2020cavity,xu2021bidirectional,fu2021cavity,xu2022light,xuLightinducedMicrowaveNoise2024}. This can be quantified by \cite{barends2011minimizing,zmuidzinas2012superconducting} \mbox{$\kappa_c^{\rm QP} = \omega_c \frac{\alpha_{\rm ki}}{\pi} \frac{n_{\rm QP}(P_l,T)}{n_0 \Delta_s} \sqrt{\frac{2\Delta_s}{\hbar \omega_c}} f(T)$}, where $\alpha_{\rm ki}$ is the kinetic inductance ratio, $n_0$ is the density of states at the Fermi energy of the superconductor, $\Delta_s$ is the superconducting gap, $n_{\rm QP}$ is the quasiparticle density which depends on pump power leakage \mbox{$P_l = \hbar \omega_p N_p \kappa^{\rm SC}_b$} and superconductor temperature $T$, and $f(T)$ is a temperature-dependant factor that approaches 1 when $k_BT \ll \hbar \omega_c$. $n_{\rm QP}$ is affected by the pump directly through the generation of quasiparticles, and indirectly by increasing the steady state superconductor temperature $T$. We calculate $T$ from a heat flow model for thin film superconductors \cite{sidorova2022phonon}, and $n_{\rm QP}$ from the quasiparticle steady state solution \cite{rothwarf1967measurement, zmuidzinas2012superconducting}, ignoring local heating in the Si resonator, expected to be less important in pulsed operation of MOCs, and which is difficult to calculate. 

Results for the quasiparticle loss rate $\kappa_c^{\rm QP}$ are shown in Fig.~\ref{fig:geo}(d) versus $N_p$ for three different microwave resonator volumes. We note that the resonator volume $V_{\rm m}$ considered in the calculation is the volume of the central line only (i.e., the inductive part where most of the quasiparticles are initially generated), as a worst case limit for the loss. The microwave cavity shows increasing quasiparticle loss with $N_p$, and cavities of smaller volume exhibit higher $\kappa_c^{\rm QP}$ at any leakage power.
These results emphasize (i) the robustness of larger microwave cavities to the optical pump and (ii) exponential reduction in microwave losses due to quasiparticle generation on the distance $d_{\rm om}$ between the optical and microwave resonators, while maintaining a high filling factor.

\subsection{Total efficiency and added noise}\label{sec:eta}
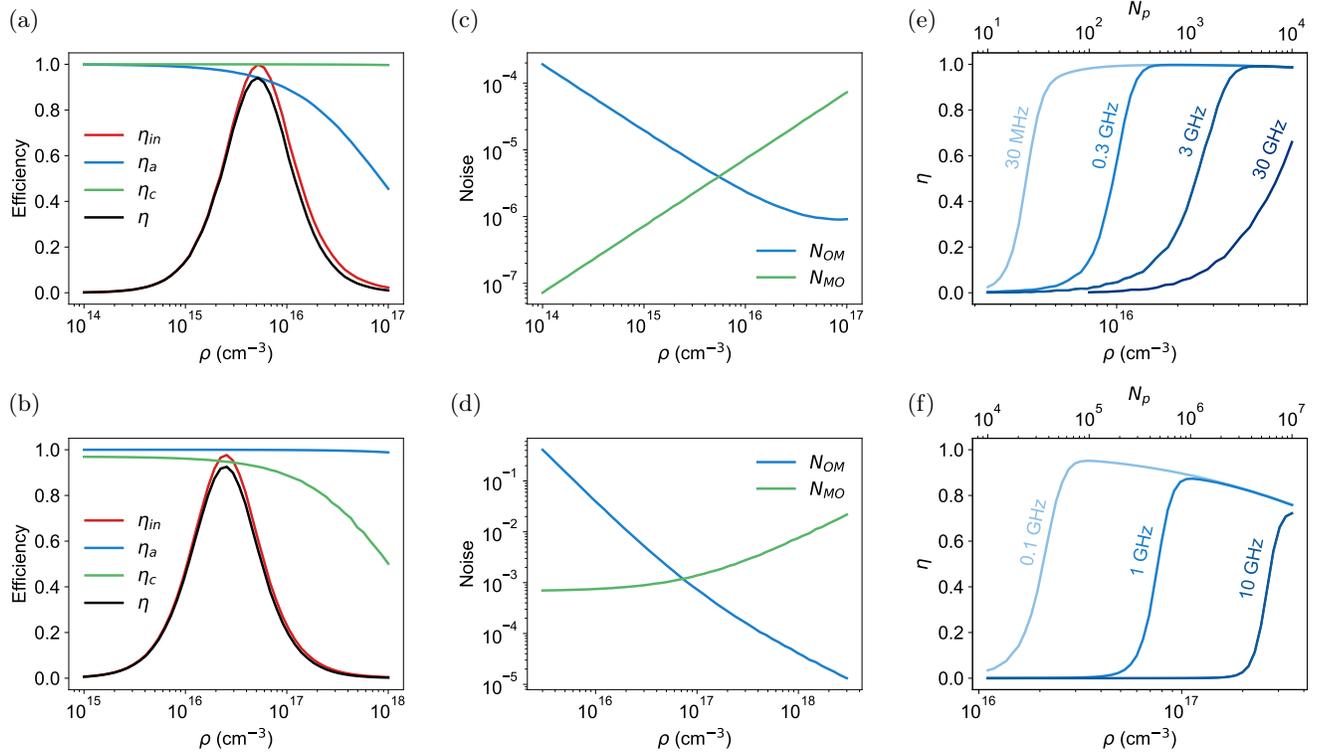
\begin{figure*}
    \centering
    \includegraphics[width=0.99\linewidth]{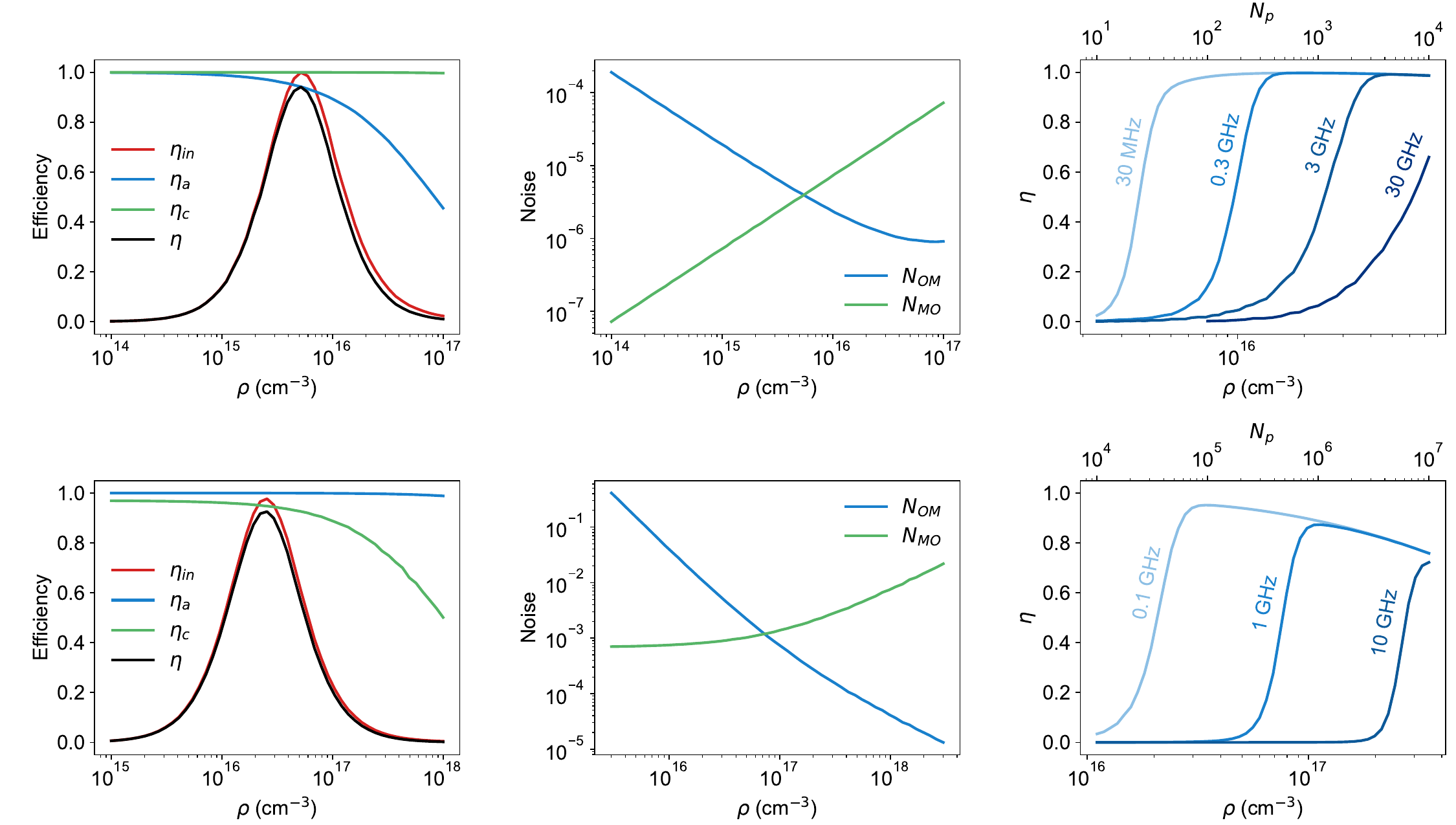}
    \put(-495,275){(a)}
    \put(-495,130){(b)}
    \put(-328,275){(c)}
    \put(-328,130){(d)}
    \put(-155,275){(e)}
    \put(-155,130){(f)}
    \caption{Efficiencies and noise for (a,c,e) T-center and (b,d,f) Er-center at zero detunings with inhomogeneous broadening (IB). (a,b) Efficiency and (c,d) noise dependence on the color center density $\rho$ at fixed number of pump photons; \mbox{$N_p=50$} in (a,c) and \mbox{$N_p=50,000$} in (b,d) for parameters in Table \ref{tab:params}. Optical IB of \qty{30}{MHz} is considered in (a,c) \cite{bergeron2020silicon} and \qty{100}{MHz} in (c,d) \cite{berkman2023millisecond}. (e,f) Optimum efficiency versus $\rho$ and $N_p$ at different values for the optical IB. At each value for $N_p$, $\rho$ is selected to give the maximum internal efficiency at zero detunings, according to the first condition in Eq.~(\ref{eqn:cond1}). Microwave IB of \qty{100}{kHz} is considered in all calculations. We consider \mbox{$V_{\rm o}=1100 \times 1.3 \times \qty{1.2}{\micro\meter^3}$} for the racetrack resonator and \mbox{$V_{\rm m}=1000 \times 2 \times \qty{0.02}{\micro\meter^3}$} for the inductive central line. We note that the phase-matching condition is satisfied for these dimensions for the TE\textsubscript{00}, TM\textsubscript{00}, and full-wave microwave modes (see the Supplementary Information).}
    \label{fig:eta}
\end{figure*}

We now predict the total quantum efficiency and added noise for our MOC operating with photons on resonance with bare microwave/optical center transitions and cavity resonances. We consider inhomogenous broadening and all sources of loss, including color center loss (Section~\ref{sec:etain}) and other losses (Section~\ref{sec:ffloss}). Microwave and optical losses are separated in the expression for total efficiency, \mbox{$\eta = \eta_{\rm in} \eta_a \eta_c$}, where \mbox{$\eta_{\rm in}=|4C|/|C+1|^2$} is the internal efficiency and \mbox{$\eta_{a}=\kappa_{a}^{\rm ex}/\kappa_{a}$} \mbox{($\eta_{c}=\kappa_{c}^{\rm ex}/\kappa_{c}$}), and $\kappa_{a}$ ($\kappa_{c}$) is now redefined as the total optical (microwave) loss including color center loss. We compute the vacuum couplings and Rabi frequency for the T-center and Er-center in Si from experimentally calibrated optical and magnetic dipoles \cite{bergeron2020silicon, gritsch2022narrow, berkman2023millisecond} and realistic cavity volumes, using \mbox{$\hbar g_{13} = d_{13} \sqrt{\frac{\hbar \omega_a F_{a}}{\epsilon_0 \epsilon_r V_{\rm o}} }$}, \mbox{$\hbar g_{12} = \mu_{12} \sqrt{\frac{\hbar \omega_c F_{c}}{V_{\rm o}/\mu_0}}$} and \mbox{$\hbar\Omega_{p} = d_{23} \sqrt{\frac{\hbar \omega_p N_p F_{b}}{\epsilon_0 \epsilon_r V_{\rm o}} }$}. Here, $d_{13}$ ($d_{23}$) is the dipole of the $\ket{1}\leftrightarrow\ket{3}$ ($\ket{2}\leftrightarrow\ket{3}$) optical signal (pump) transition, $\mu_{12}$ is the magnetic dipole of the $\ket1\leftrightarrow\ket2$ microwave transition, $V_{\rm o}$ is the volume of the nonlinear medium, $F_{i }$ is the unitless filling factor of mode $i\in \{a,b,c\}$, and $F_{abc}=\sqrt{F_a F_b F_c}$ for spatially uniform bosonic fields. We consider cavity coupling losses of \mbox{$\kappa_a^{\rm ex}=\qty{2}{GHz}$ ($Q_a^{\rm ex} \sim 10^5$)} and \mbox{$\kappa_c^{\rm ex}=\qty{0.8}{MHz}$ ($Q_c^{\rm ex} \sim 10^4$)}. For the T-center, we use $N_p=50$, while for the Er-center we use $N_p=50,000$ to compensate for the relatively small optical dipole of the Er-center. 

Results are presented in Fig.~\ref{fig:eta} for T-centers and Er-centers versus the center density $\rho=N_A/V_{\rm o}$ using color center parameters from Table~\ref{tab:params}, illustrating the potential of our proposed approach for efficient transduction. In both cases we predict efficiency as high as $95\%$ with the number of noise photons close to zero.

We find the efficiency of the T-center device is limited by the optical external efficiency, $\eta_a$ (Fig.~\ref{fig:eta}(a)), while for the Er-center, the efficiency is limited by the microwave external efficiency, $\eta_c$ (Fig.~\ref{fig:eta}(b)). 
This is because the relatively strong optical dipole (magnetic dipole) of the T-center (Er-center) results in a large center-induced optical (microwave) loss, which increases with the center density.
One way to alleviate the effect of the center-induced loss in one of the cavities is to increase its reservoir coupling and maintain the matching condition by decreasing the reservoir coupling of the other cavity.
Hence, color centers with strong optical dipoles, such as the T-center, require an optical cavity with a relatively small $Q^{\rm ex}_a$ and a microwave cavity with a large $Q^{\rm ex}_c$, while the opposite is true for color centers with strong microwave dipoles, such as the Er-center.

Results for added noise (Eq.~(\ref{eqn:nois})) are presented in Fig.~\ref{fig:eta}(c,d). The microwave-optical noise, $N_{\rm MO}$, increases with density due to increasing the center-induced microwave loss. In the case of Er in Fig.~\ref{fig:eta}(d), $N_{\rm MO}$ is independent of the density at small densities, because the microwave intrinsic loss is dominated by the pump-induced loss rather than the center-induced loss. For the optical-microwave noise, $N_{\rm OM}$, the increase of the density has two effects. First, it increases the microwave intrinsic loss, which increases the noise linearly, assuming an over-coupled optical cavity. Second, it increases $\xi_{ac}$ linearly, which reduces the noise quadratically, giving an overall linear decrease in the noise with increasing  density. 
The two noise lines intersect when matching ($C=1$) is obtained, as expected from Eq.~(\ref{eqn:nois}) for over-coupled cavities. Importantly, for both the Er- and T-centers, the calculated noise at the optimum density is significantly less than the threshold, 0.5, for the quantum state transfer by direct conversion \cite{wolf2007quantum}.

We study the effect of the inhomogeneous broadening on the conversion efficiency in Fig.~\ref{fig:eta}(e,f). The total efficiency is plotted for different optical inhomogeneous broadenings versus the center density and the number of pump photons, which are varied together to keep the condition $C=1$ satisfied at all points. As the inhomogeneous broadening increases, a larger number of pump photons and higher center density are required to realize the near-unity efficiency. If the number of pump photons is too large, the pump-induced microwave loss lowers the microwave external efficiency, as shown in Fig.~\ref{fig:eta}(f). Achieving high efficiency using our MOC relies on the realization of small inhomogeneous broadening for a practical center density.

Several practical challenges present themselves in real devices. For example, the narrow inhomogeneous broadening of the T-center that has been measured for isotope-purified bulk Si \cite{bergeron2020silicon} has yet to be demonstrated in Si integrated optical cavities. Another potential challenge is integrating a superconducting resonator with $Q\sim 50,000$ on a silicon-on-insulator platform. However, $Q_i>10^5$ has recently been realized for suspended resonators using new fabrication approaches \cite{fang2020suspended}. Furthermore, employing suitable techniques for tuning the optical modes will be necessary for frequency matching the color center transitions and for fulfilling the triple-resonance condition with the microwave mode. 
Color centres with multiple resonance lines represent a challenge. For both T- and Er-centers, multiple resonance lines appear in magnetic field \cite{bergeron2020silicon,berkman2023observing,rinner2023erbium}. Additionally, Er can show multiple resonance lines at zero magnetic field \cite{przybylinska1996optically,kenyon2005erbium,berkman2023millisecond}; however the number of these states can be significantly reduced by careful sample preparation \cite{gritsch2022narrow}.
All these aspects require careful design and fabrication of the device for experimental realization.

\section{Conclusion}
We have developed a theory of microwave-optical conversion for arbitrary bare microwave/optical resonator frequencies relative to the bare microwave/optical transitions of an ensemble of color centers, including on-resonance where strong ensemble-cavity coupling effects are obtained. We find high-efficiency operating points for microwave and optical signal photon frequencies that are nearly resonant with hybrid center-cavity resonances, and a counterintuitive operating point where microwave and optical signal photons are nearly resonant with the bare center-cavity frequencies, but far from the hybrid center-cavity resonances. The strong coupling regime offers significantly enhanced nonlinearity at lower pump powers for conversion, and the counter-intuitive operating point offers enhanced robustness to inhomogeneous broadening. Accounting for inhomogeneous broadening, color center-induced loss, Si two-photon and free-carrier absorption, and optical and microwave mode losses due to optically generated quasiparticles, we find that high conversion efficiency ($\sim 95$~\%) and low noise ($\ll 1$~photon) is possible for practically obtainable optical and microwave intrinsic quality factors of $\sim10^6$ and $\sim5\times 10^4$ respectively, and around $\sim\qty{1}{GHz}$ of inhomogeneous broadening.

\begin{acknowledgments}
 The authors acknowledge Lukas Chrostowski, Marek Korkusinski and Shabir Barzanjeh for helpful discussions and  Alexey Lyasota and Sven Rogge for pointing out new work on Er doped Si. This work was undertaken with support from the Stewart Blusson Quantum Matter Institute (SBQMI), and the Canada First Research Excellence Fund, Quantum Materials and Future Technologies Program. JS and JFY acknowledge financial support from the National Science and Research Council of Canada (NSERC) in the Discovery Grant program and Quantum Alliance Consortium ``CanQuEST'', from the National Research Council in the Quantum Sensors Challenge program, and from Defence Canada (Innovation for Defence Excellence and Security, IDEaS). JS and JFY acknowledge financial support from the Canadian Foundation for Innovation (CFI) John Edwards Leaders Foundation and CFI Innovation Fund. MK acknowledges financial support from the SBQMI QuEST fellowship program. PSK acknowledges financial support from the NSERC CREATE in Quantum Computing Program (Grant Number 543245). The authors acknowledge CMC Microsystems for the provision of computer aided design tools that were essential to obtain the results presented here. This research was supported in part through computational resources provided by Advanced Research Computing at the University of British Columbia.
\end{acknowledgments}

% \appendix
% \section{Appendixes}

%\nocite{*}

%\bibliography{refs}% Produces the bibliography via BibTeX.

\end{document}